\documentclass{elektr}
\usepackage{hyperref}
\usepackage{subfig}
\usepackage{float}
\hypersetup{
colorlinks=true,
urlcolor=blue,
citecolor=blue}
\usepackage[all]{xy,xypic}
\usepackage{amsfonts,amssymb,amsmath,amsgen,amsopn,amsbsy,theorem,graphicx,epsfig}
\usepackage{eufrak,amscd,bezier,latexsym,mathrsfs,eurosym,enumerate}
\usepackage[utf8]{inputenc}
\usepackage[english]{babel}
\usepackage{cleveref,multirow}
\usepackage[dvipsnames]{xcolor}
\usepackage[pagewise]{lineno}

\yil{}
\vol{}
\fpage{}
\lpage{}
\doi{}

\title{Design and Modeling of a PVDF-TrFe Flexible Wind Energy Harvester}

\author[KULLUKÇU et al.]{
\textbf{Berkay KULLUKÇU$^{1}$~, Levent BEKER$^{2}$\thanks{lbeker@ku.edu.tr}~}\\
$^{1}$Mechanical Engineering, Graduate School of Science and Engineering, Koç University, İstanbul, Turkey, \\ ORCID iD: https://orcid.org/
0000-0002-0784-1537\\
$^{2}$Mechanical Engineering, Graduate School of Science and Engineering, Koç University,\\ İstanbul, Turkey, ORCID iD: https://orcid.org/0000-0002-9777-6619
\\
[1.8em]

\rec{.201}
\acc{.201}
\finv{..201}
}

\def\E{\ifmmode{\mathbb E}\else{$\mathbb E$}\fi} %natural numbers
\def\N{\ifmmode{\mathbb N}\else{$\mathbb N$}\fi} %natural numbers
\def\R{\ifmmode{\mathbb R}\else{$\mathbb R$}\fi} %real numbers
\def\Q{\ifmmode{\mathbb Q}\else{$\mathbb Q$}\fi} %rational numbers
\def\C{\ifmmode{\mathbb C}\else{$\mathbb C$}\fi} %complex numbers
\def\H{\ifmmode{\mathbb H}\else{$\mathbb H$}\fi} %complex numbers
\def\Z{\ifmmode{\mathbb Z}\else{$\mathbb Z$}\fi} %integers
\def\P{\ifmmode{\mathbb P}\else{$\mathbb P$}\fi} %real numbers
\def\T{\ifmmode{\mathbb T}\else{$\mathbb T$}\fi} %real numbers
\def\SS{\ifmmode{\mathbb S}\else{$\mathbb S$}\fi} %real numbers
\def\DD{\ifmmode{\mathbb D}\else{$\mathbb D$}\fi} %real numbers

\newcommand{\bse}{\begin{subequations}}
\newcommand{\ese}{\end{subequations}}
\newcommand{\ben}{\begin{enumerate}}
\newcommand{\een}{\end{enumerate}}
\newcommand{\bens}{\begin{enumerate*}}
\newcommand{\eens}{\end{enumerate*}}
\newcommand{\be}{\begin{equation}}
\newcommand{\ee}{\end{equation}}
\newcommand{\bea}{\begin{eqnarray}}
\newcommand{\eea}{\end{eqnarray}}
\newcommand{\baa}{\begin{eqnarray*}}
\newcommand{\eaa}{\end{eqnarray*}}
\newcommand{\bc}{\begin{center}}
\newcommand{\ec}{\end{center}}

\newcommand{\vs}{\vspace}

\theoremstyle{corollary}

\theoremstyle{lemma}

\theoremstyle{proposition}

\theoremstyle{axiom}

\theoremstyle{conjecture}

\theoremstyle{example}

\theoremstyle{definition}
\theoremstyle{remark}
\begin{document}

\maketitle

\begin{abstract}This study presents the simulation, experimentation, and design considerations of a Poly(vinylidene fluoride co-trifluoroethylene)/ Polyethylene Terephthalate (PVDF-TrFe / PET), laser-cut, flexible piezoelectric energy harvester. It is possible to obtain energy from the environment around autonomous sensor systems, which can then be used to power various equipment. This article investigates the actuation means of ambient vibration, which is a good candidate for using piezoelectric energy harvester (PEH) devices. The output voltage characteristics were analyzed in a wind test apparatus. Finite element modeling (FEM) was done for von Mises stress and modal analysis. Resonance frequency sweeps, quality factors, and damping ratios of the circular plate were given numerically. For a PVDF-TrFe piezoelectric layer thickness of 18 µm and 1.5 mm radius, a damping ratio of 0.117 and a quality factor of 4.284 was calculated. $V_{max}$ was calculated as 984 mV from the wind setup and compared with the FEM outputs.

\end{abstract}

\section{Introduction}
\label{Int}

Microelectronics’ inclusion and rapid development have profoundly impacted human
society in recent decades. The widespread availability of intelligent electronic devices paved a way for a surge of varied applications that may improve one’s quality of
life in a variety of diﬀerent ways. Concurrent advancements in microelectromechanical systems (MEMS) technologies have resulted in improved device performance
with reductions in size, cost, and the amount of power they use. Together with
advancements in the capacity of wireless transmission, the tendency toward devices
that are both smaller and inexpensive makes it feasible to establish distributed networks. Such networks have the potential to serve a variety of functions. Observing
temperature, light, and the position of individuals in commercial properties to regulate the environment in a manner that is more energy-efficient, detecting dangerous
chemical agents in high-density areas, monitoring the formation of fatigue cracks on
aircraft, monitoring the speed and pressure in vehicle tires, and a great deal more are
all examples of the many applications of this technology. Numerous experts believe
that very low-power embedded electronic devices will become ubiquitous, performing
tasks ranging from factory automation to entertainment. However, the performance
of such new advancements is reliant on the availability of reliable sources of energy in
the near future. As Batteries and other types of energy sources that have a set capacity
present a number of challenges due to their cost, size, and longevity, the device presented in the article, is intended to provide the necessary power for electronic devices by harvesting vibrational energy. Table (\ref{tab1}) shows the comparison of macro and micro-scale energy harvesters from literature.

\iffalse
\begin{figure}[H]
\begin{center}
\includegraphics[width=9.5cm]{Picture1.png}
\caption{3D representation of the energy harvester.}
\label{fig1}
\end{center}\vs{-4mm}
\end{figure}
\fi
\begin{table}[H]
\caption{Energy Harvester Parameters Compared.}
\begin{center}
\begin{tabular}{|l|l|l|l|l|l|}
\hline
Paper &
  Dimensions &
  Shape &
  Voltage Output &
  \begin{tabular}[c]{@{}l@{}}Power \\ Output\end{tabular} &
  \begin{tabular}[c]{@{}l@{}}Resonance\\ Frequency\end{tabular} \\ \hline
{[}1{]} Qian, F. et. al. &
  \begin{tabular}[c]{@{}l@{}}150x19x3 \\ mm for single \\ beam\end{tabular} &
  \begin{tabular}[c]{@{}l@{}}Venus \\ flytrap \\ shape\end{tabular} &
  \begin{tabular}[c]{@{}l@{}}Reported \\ in figures\end{tabular} &
  0.193 mW &
  78.35 rad/s \\ \hline
{[}2{]} Li, X. et. al. &
  110x22 mm &
  \begin{tabular}[c]{@{}l@{}}Sandwich \\ structure\end{tabular} &
  \begin{tabular}[c]{@{}l@{}}6.21 V \\ max. at a \\ resonance\\peak\end{tabular} &
  \begin{tabular}[c]{@{}l@{}}Reported\\in\\figures\end{tabular} &
  \begin{tabular}[c]{@{}l@{}}18.18 Hz,\\24.74 Hz,\\and 28.12\\Hz \end{tabular} \\ \hline
{[}3{]} Ramírez, J. M. et. al. &
  \begin{tabular}[c]{@{}l@{}}Many\\different\\ beam lengths,\\btw. 85 to 150\\mm\end{tabular} &
  \begin{tabular}[c]{@{}l@{}}Linked\\E-shape\\multi-beam\end{tabular} &
  6 V max. &
  200 $\mu$W &
  4.7 Hz \\ \hline
{[}5{]} Morimoto et. al. &
  \begin{tabular}[c]{@{}l@{}}18.5x5 \(mm^2\) \\ cantilever \end{tabular} &
  \begin{tabular}[c]{@{}l@{}}Cantilever\end{tabular} &
  \begin{tabular}[c]{@{}l@{}}2.6 V\end{tabular} &
  244 $\mu$W &
  126 Hz \\ \hline
{[}6{]} Horowitz et. al. &
  \begin{tabular}[c]{@{}l@{}}Reported in\\figures for\\different\\designs \end{tabular} &
  \begin{tabular}[c]{@{}l@{}}Circular\\diaphragm\end{tabular} &
  \begin{tabular}[c]{@{}l@{}}4.6 mV\end{tabular} &
  \begin{tabular}[c]{@{}l@{}}0.34 \(\mu W/cm^2\)\\power\\density\end{tabular} &
  3.7 kHz \\ \hline
{[}7{]} Wang et. al. &
  $\sim$2\(mm^2\) &
  Nanowire &
  \begin{tabular}[c]{@{}l@{}}Reported \\ in figures\\(mV range)\end{tabular} &
  \begin{tabular}[c]{@{}l@{}}Reported\\in figures\end{tabular} & 
  $\sim$3 MHz \\ \hline
{[}8{]} Hu et. al. &
  1.5×0.5 \(cm^2\) &
  Nanowire &
  1.5 V &
  \begin{tabular}[c]{@{}l@{}}70\\ \(\mu W/cm^3\)\\power\\density\end{tabular} &
  Not given \\ \hline
{[}10{]} Lin et. al. &
  \begin{tabular}[c]{@{}l@{}}500 nm x 6 \\$\mu$m nanowire\end{tabular} &
  Nanowire &
  8V &
  \begin{tabular}[c]{@{}l@{}}$\sim$5.3 \\mW\(/cm^3\)\end{tabular} &
  Not given \\ \hline
\end{tabular}
\label{tab1}
\end{center}\vs{-4mm}
\end{table}
\section{PVDF-TrFe Sensor Modeling}
\subsection{Material Search}
The polymer PVDF-TrFe is selected for this application for various purposes. First of all, it can be stretched in both the longitudinal and the transverse
directions, which allows it to be used to form a bi-oriented film [11]. If PVDF
proves to be an excellent candidate for the fabrication of piezo films, the P(VDFTrFE) copolymer may also be poled [11]. This particular copolymer will crystallize
immediately in beta form as a result of the replacement of a few of the VF2(CH2-
CF2) molecules in the PVDF with VF3 (CH-CF3) molecules [11].
In the study of Mohammadi et al., Fourier transforms infrared spectroscopy was
used to evaluate the inﬂuence of biaxial orientation and deformation rate on the
microstructure of stretched PVDF samples, and wide-angle X-ray diﬀraction was
used to confirm the findings of the FTIR analysis [12]. FTIR spectroscopy is a
well-established method for determining whether or not there has been a shift in the amount of b-phase present in PVDF films. It has been demonstrated that an
additional transition of the non-polar a phase into the polar b crystallites occurs
when the stretching rate is 10–50 cm/min [12]. Polarized PVDF flms have fairly
balanced piezoelectric activity in the flm plane, unlike uniaxially oriented PVDF
flms, which have more signifcant piezoelectric coefcients [13]. This is because
polarized PVDF flms are oriented in two orthogonal directions. The crystalline
structure of PVDF is directly responsible for the material’s electrical characteristics
[14-18]. Today, various diﬀerent polar crystals have been described for PVDF. These
modifcations are designated as b (form I), g (form III), d (form IV), and e (form
V) [19]. Only when the stretch ratio is more than five is the melt-crystallized
PVDF capable of being transformed into the b-phase [20]. In order to produce
polar b-phase PVDF straightforwardly, one must either cast PVDF from solutions
of hexamethyl phosphoric triamide or quench molten PVDF under high pressure.
Both of these processes are required. Both of these methods produce the same result
[20]. The electrical charge in the matching PVDF film is significantly lower than
that in the 55 mol percent VDF copolymer, which has the biggest dielectric constant
and the most considerable electric current [12]. An uncommon negative longitudinal piezoelectric eﬀect was seen by Katsouras
et al. in the ferroelectric polymer poly(vinylidene ﬂuoride) (PVDF), as well as its
copolymers with triﬂuoroethylene (P(VDF-TrFE)) [21]. It seems inconceivable that
the polarization-biased electrostrictive contribution of the crystalline component is
the only factor responsible for this event [21]. Electromechanical contact between
crystalline lamellae and amorphous areas is the cause [105]. The so-called dimensional model, a commonly accepted explanation for the negative piezoelectric eﬀect
of PVDF, assumes that the dipoles are stiﬀ and maintain their fixed moment and
orientation even when the material is mechanically deformed [21].
According to Katsouras et al., applying a positive stress perpendicular to the
chains mainly increases the distance between the chains. This is because the interactions between the chains, both van der Waals and electrostatic, are relatively
weak. This dipole-induced piezoelectricity process has been verified using quantum mechanical simulations for single-crystalline polymers [22, 23]. P(VDF-TrFE)
crystallizes into needle shaped domains, whereas PVDF films consist of massive
lamellae, which spread outward from the center. Table (\ref{tab2}) shows the superiority of PVDF-TrFE compared with other widely used piezoelectric materials in terms of voltage output KPI.
\begin{table}[H]
\caption{Piezoelectric Material Parameters Compared.}
\begin{center}
\begin{tabular}{|l|l|l|l|l|l|}
\hline
Parameter &
  $\text{BaTiO}_{\text{3}}$ {[}26,28{]}&
  Quartz {[}30,31{]}&
  PZT4 {[}25,30,32{]} &
  PZT5H {[}24,27,30{]}&
  PVDF-TrFe {[}25,29{]}\\
 \hline
 $g_{33}$ {[}\(10^{-3}\){]} &
  13.06 &
  51.98 &
  25.12 &
  19.71 &
  456.79 \\ \hline
\end{tabular}
\label{tab2}
\end{center}\vs{-4mm}
\end{table}
As given in (\ref{eq1}) and (\ref{eq2}), the $g_{33}$ constant is directly proportional to the voltage output of the piezoelectric material. PVDF-TrFe has one order of magnitude higher $g_{33}$ constant than most piezoelectric materials.

\begin{equation}
\label{eq1}
V = g_{33}ds
\end{equation}
\begin{equation}
\label{eq2}
V = g_{33}dE\varepsilon
\end{equation}
\subsection{Proposed Design}
\begin{figure}[H]
\begin{center}
    \centering
    \subfloat[\centering]{{\includegraphics[width=7.5cm]{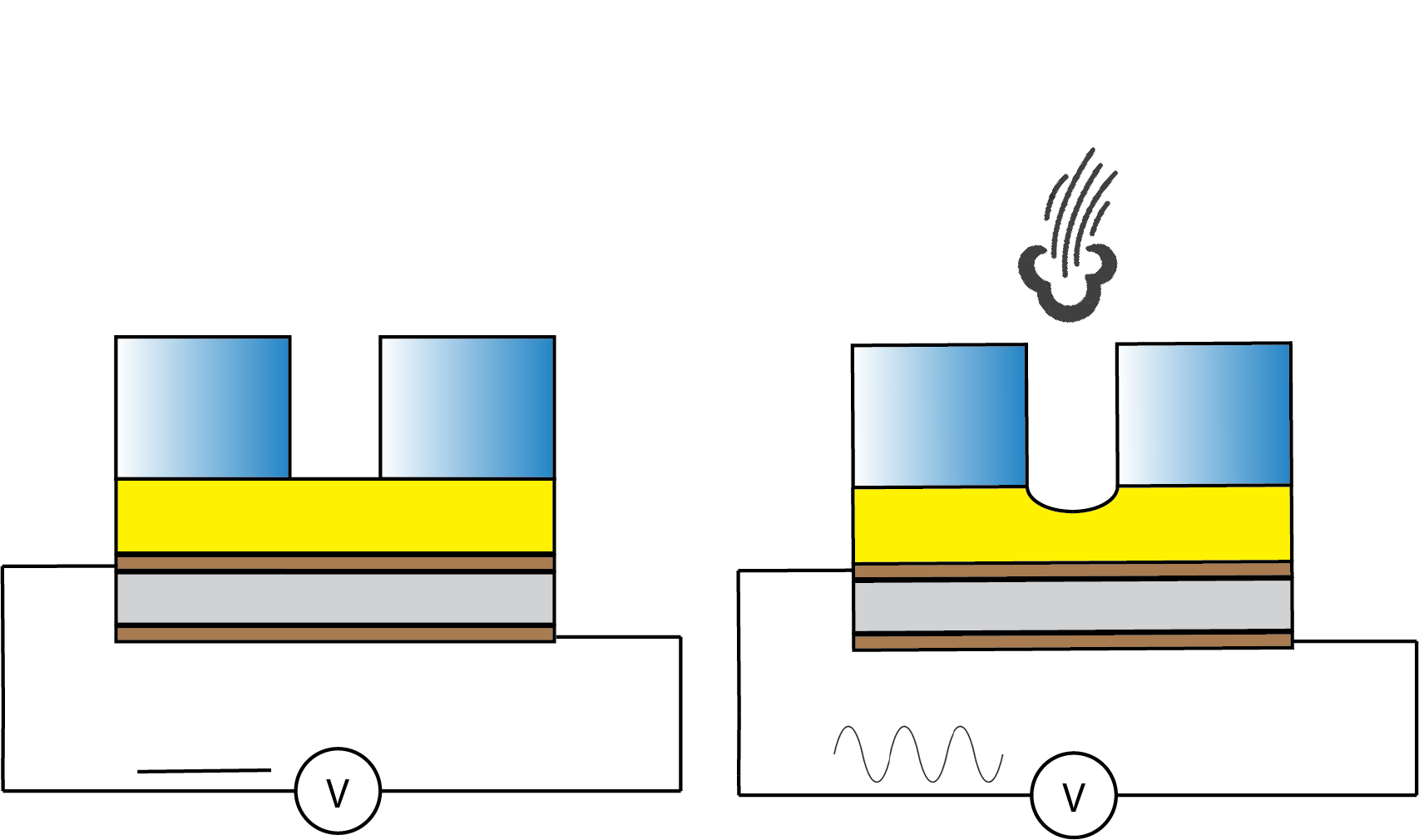} }}%
    \qquad
    \subfloat[\centering]{{\includegraphics[width=6cm]{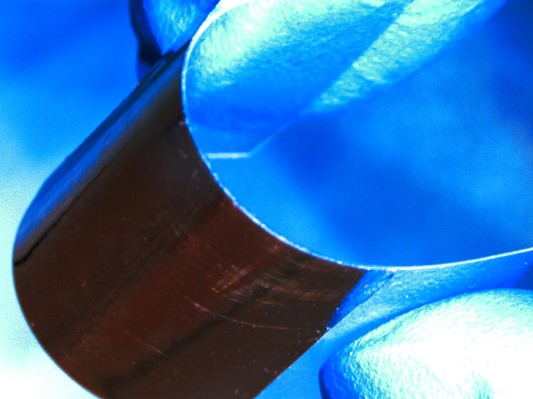} }}%
    \caption{Working principle of the fabricated harvester: a) Circular plates vibrate under wind pressures b) The fabricated device under an optical microscope.}\label{fig2}
    \end{center}\vs{-4mm}
\end{figure}

The device operates under wind vibration. Figure (\ref{fig2}) presents an illustration of the device's functionality and an image taken under optical microscope. Wind vibration deforms the piezoelectric plates and creates AC voltage and current. The power harvester can power up low-power electronics or charge electrical storage devices.

To fabricate the device, plain PET sheets were cut to 10 mm by 10 mm using a mechanical cutter. The sheets had a thickness of 140 $\mu$m and were laser-cut to create 3 mm in diameter circular perforations. The PET was then covered with a double-sided tape measuring 65 microns in thickness, and PolyK PVDF film adhered to the sheet. In addition, the tape served as the harvester's structural layer, shifting the piezoelectric layers away from the neutral axis.

\subsection{Sensor Modeling}
\begin{figure}[H]
\begin{center}
    \centering
    \subfloat[\centering]{{\includegraphics[width=8cm]{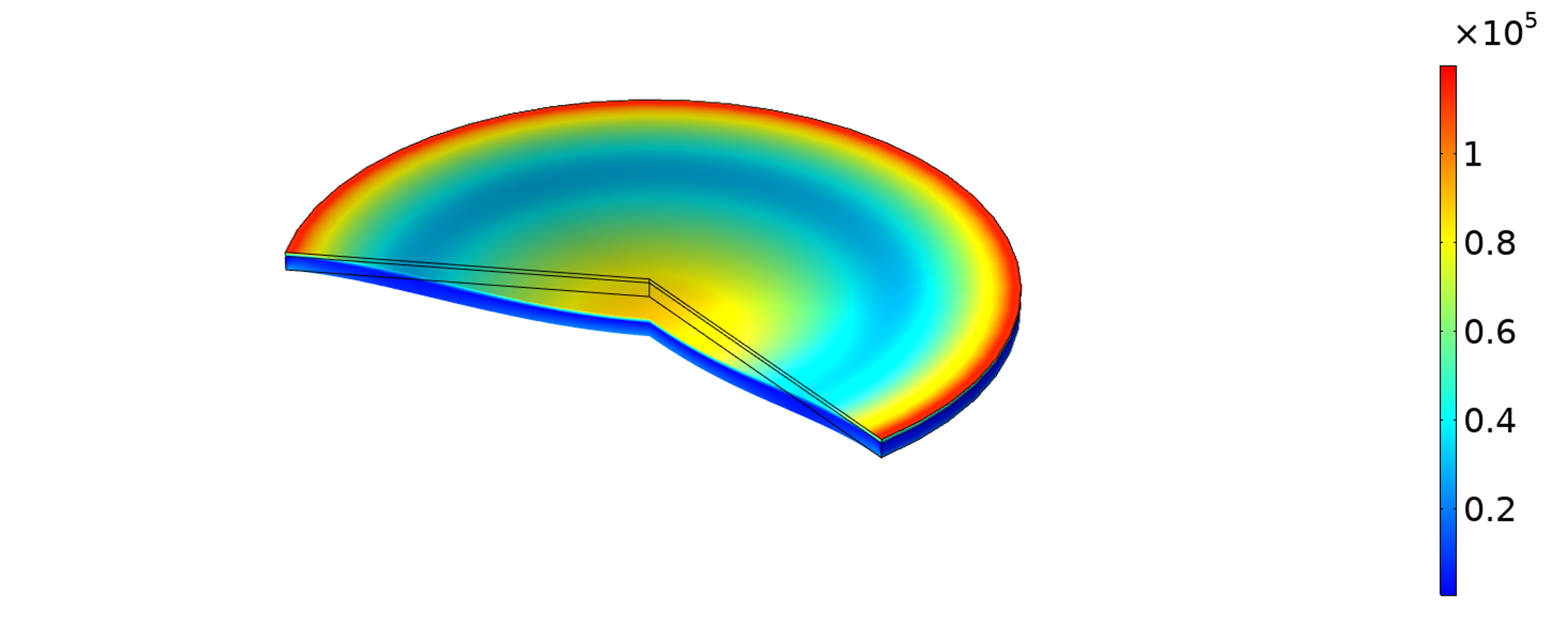} }}%
    \qquad
    \subfloat[\centering]{{\includegraphics[width=8cm]{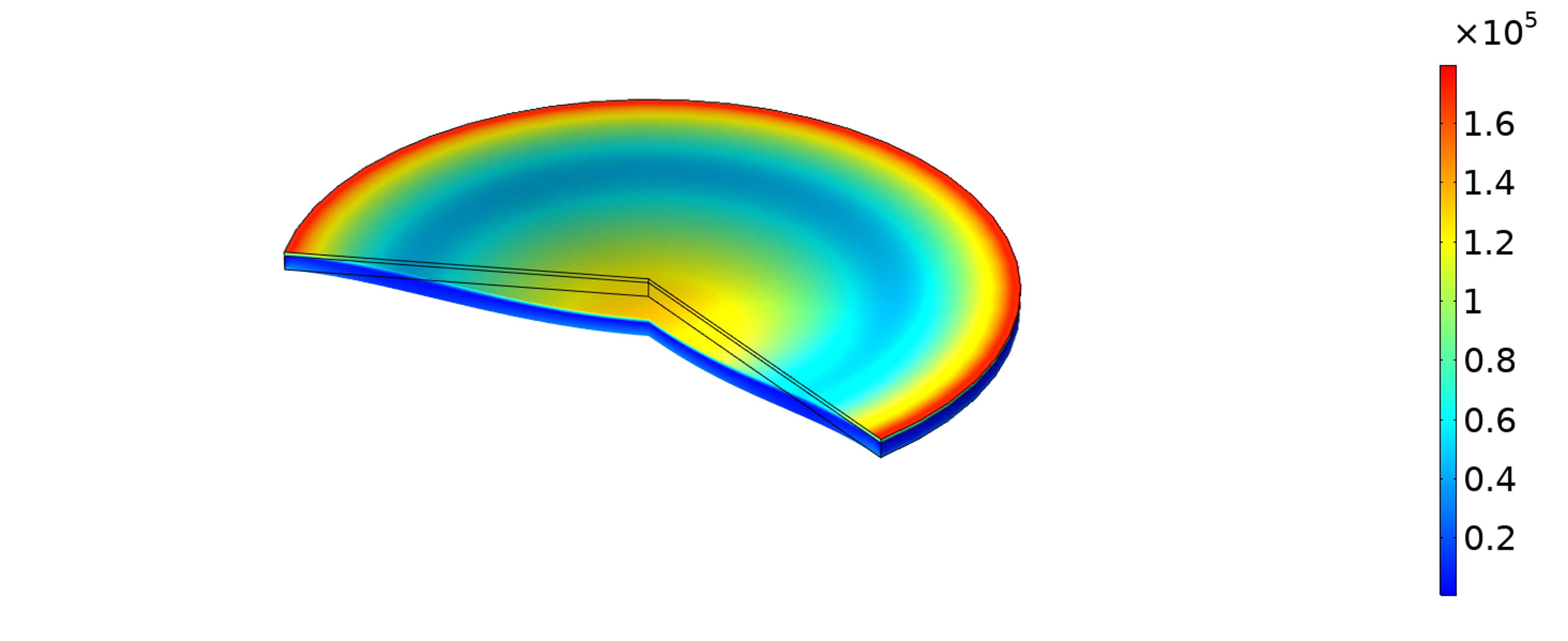} }}%
    \vskip\baselineskip
    \centering
    \subfloat[\centering]{{\includegraphics[width=8cm]{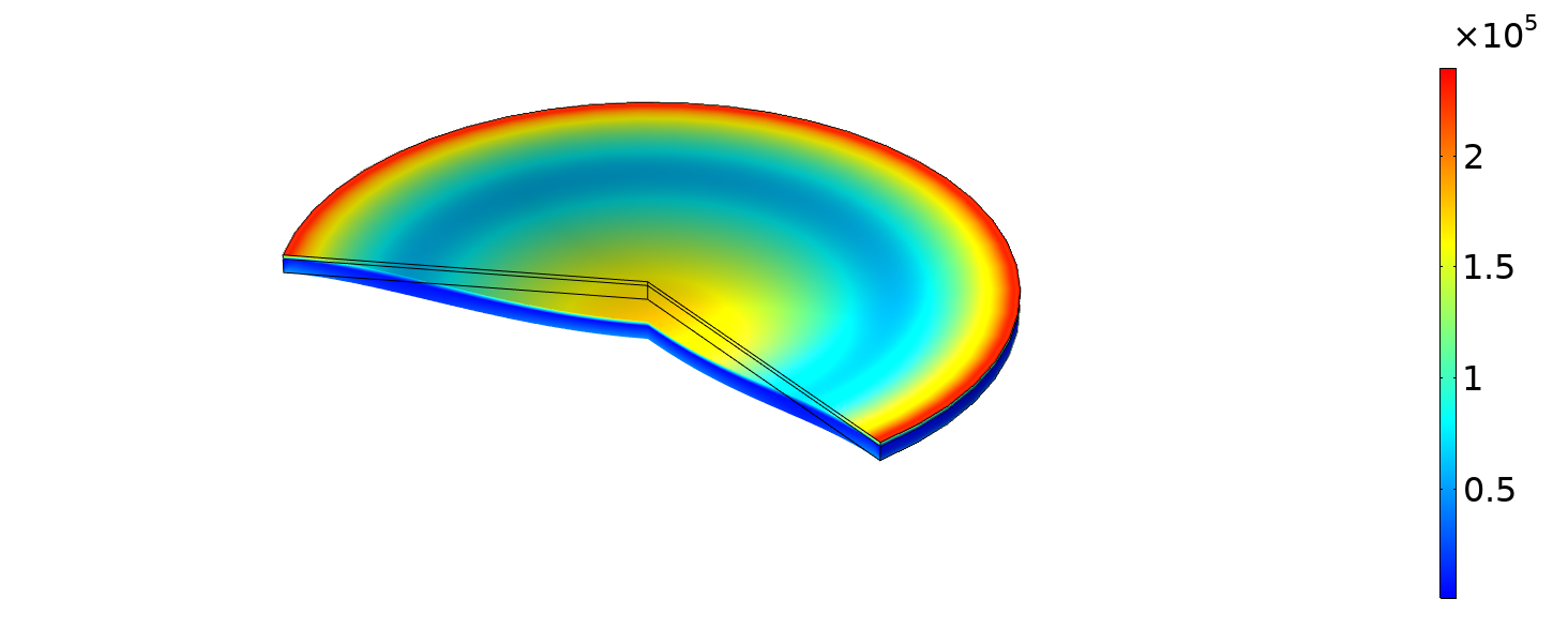} }}%
    \qquad
    \subfloat[\centering]{{\includegraphics[width=8cm]{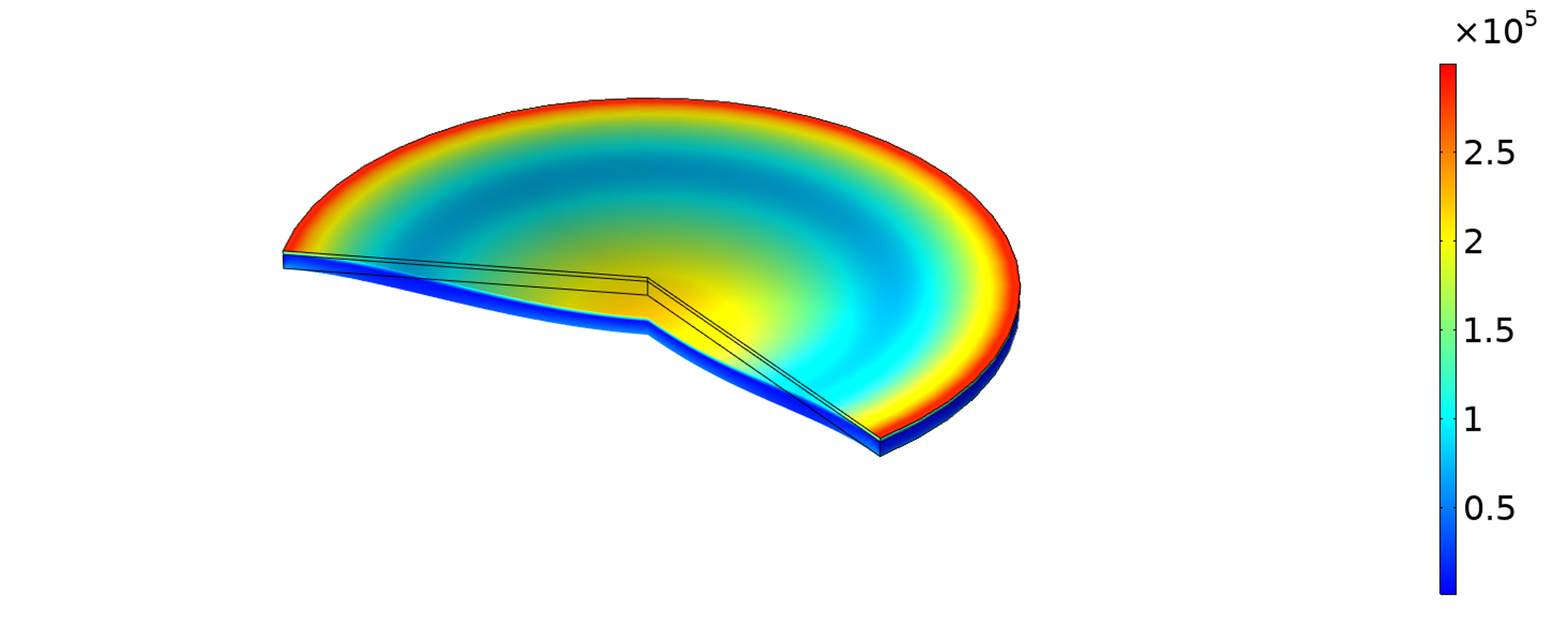} }}%
    \caption{Von Mises stress analysis of the circular plate for different wind pressure
inputs: a) 200 Pa b) 300 Pa c) 400 Pa d) 500 Pa pressure.}\label{fig:fig4}
\end{center}\vs{-4mm}
\end{figure}

\begin{figure}[H]
\begin{center}
    \centering
    \subfloat[\centering]{{\includegraphics[width=8cm]{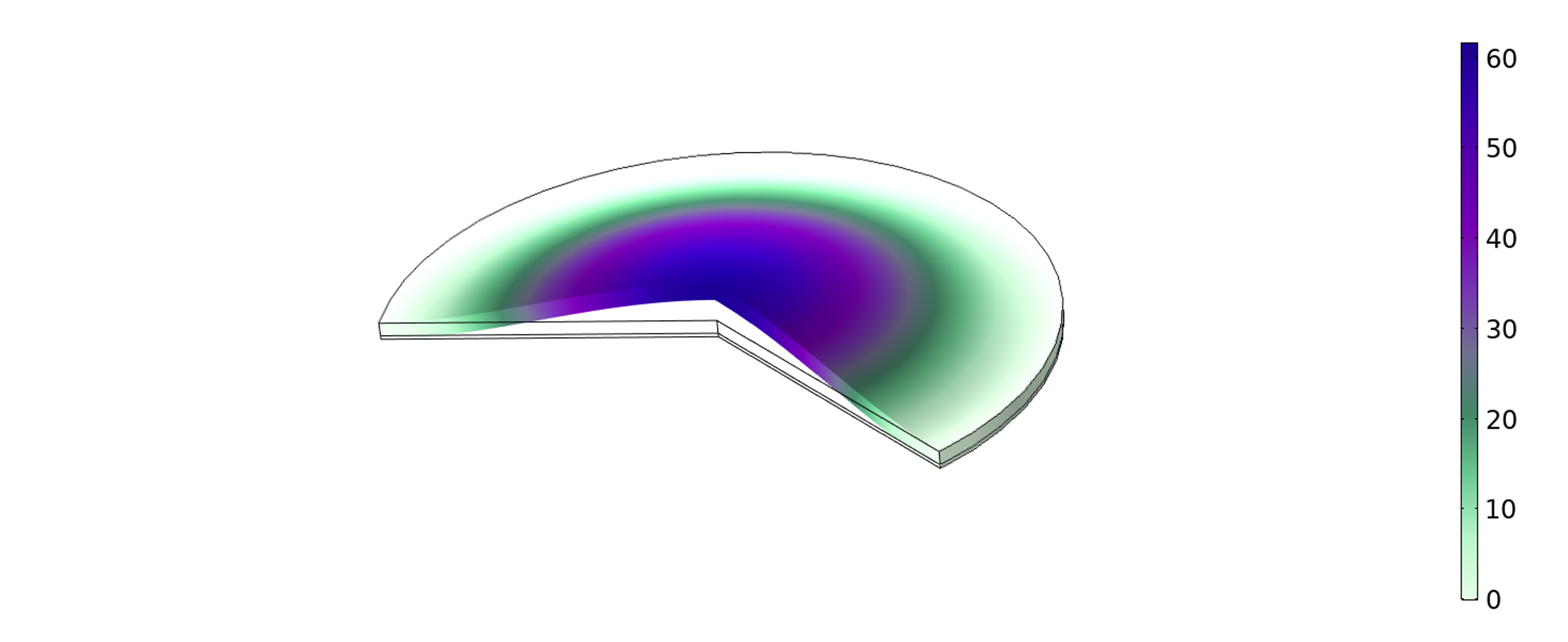} }}%
    \qquad
    \subfloat[\centering]{{\includegraphics[width=8cm]{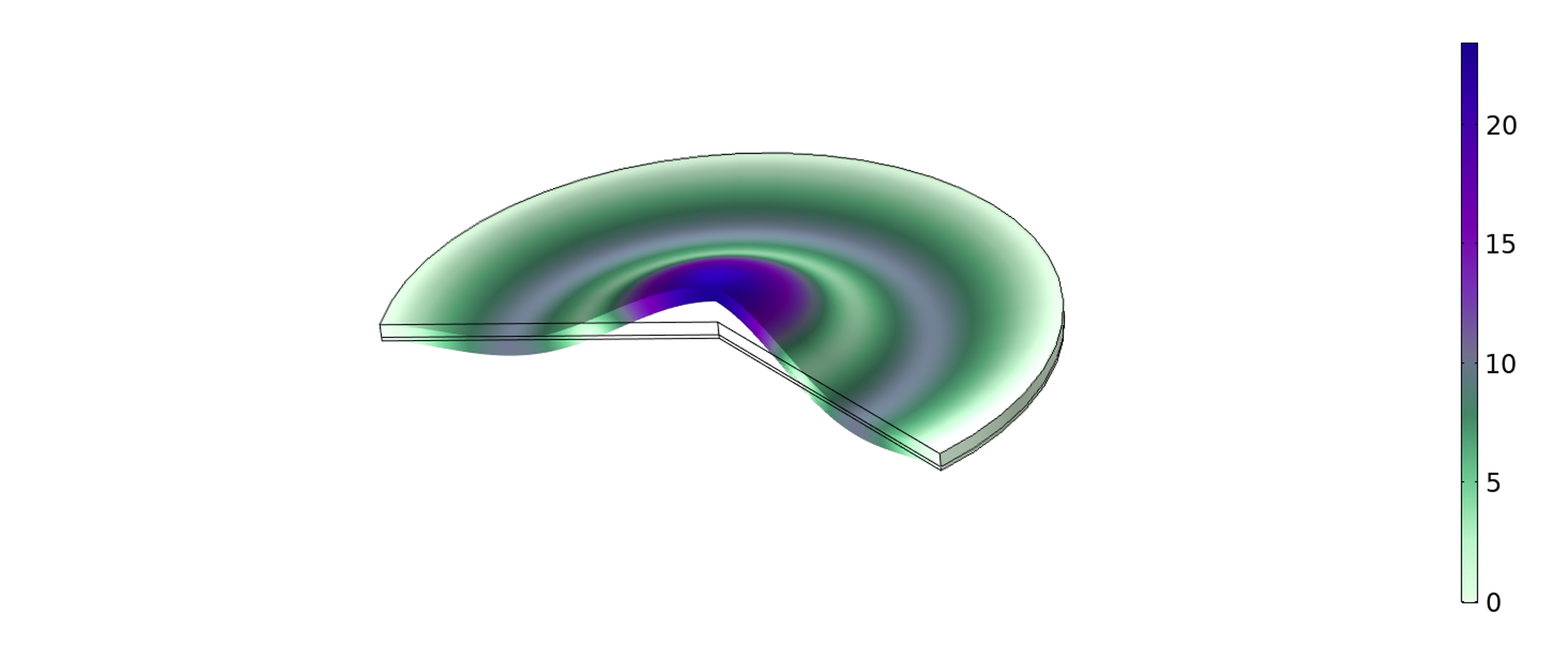} }}%
    \vskip\baselineskip
    \centering
    \subfloat[\centering]{{\includegraphics[width=8cm]{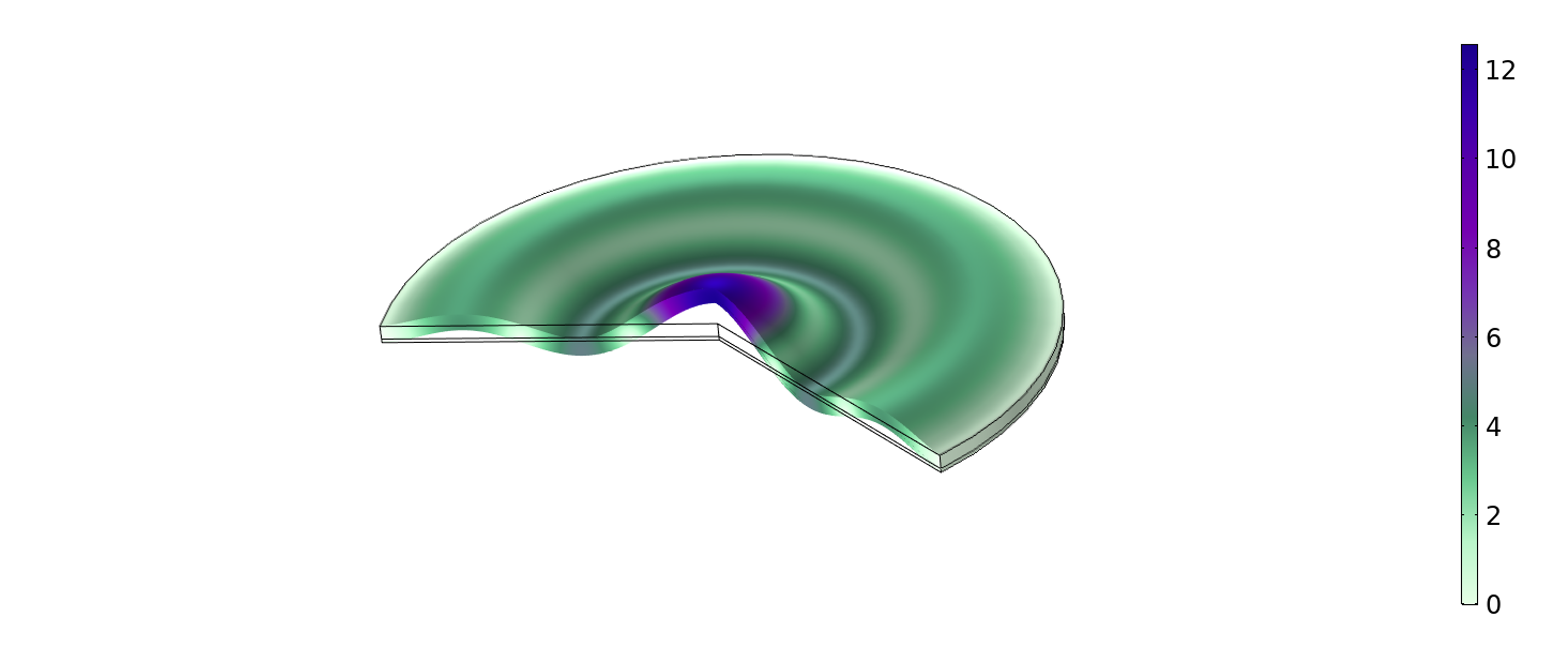} }}%
    \qquad
    \subfloat[\centering]{{\includegraphics[width=8cm]{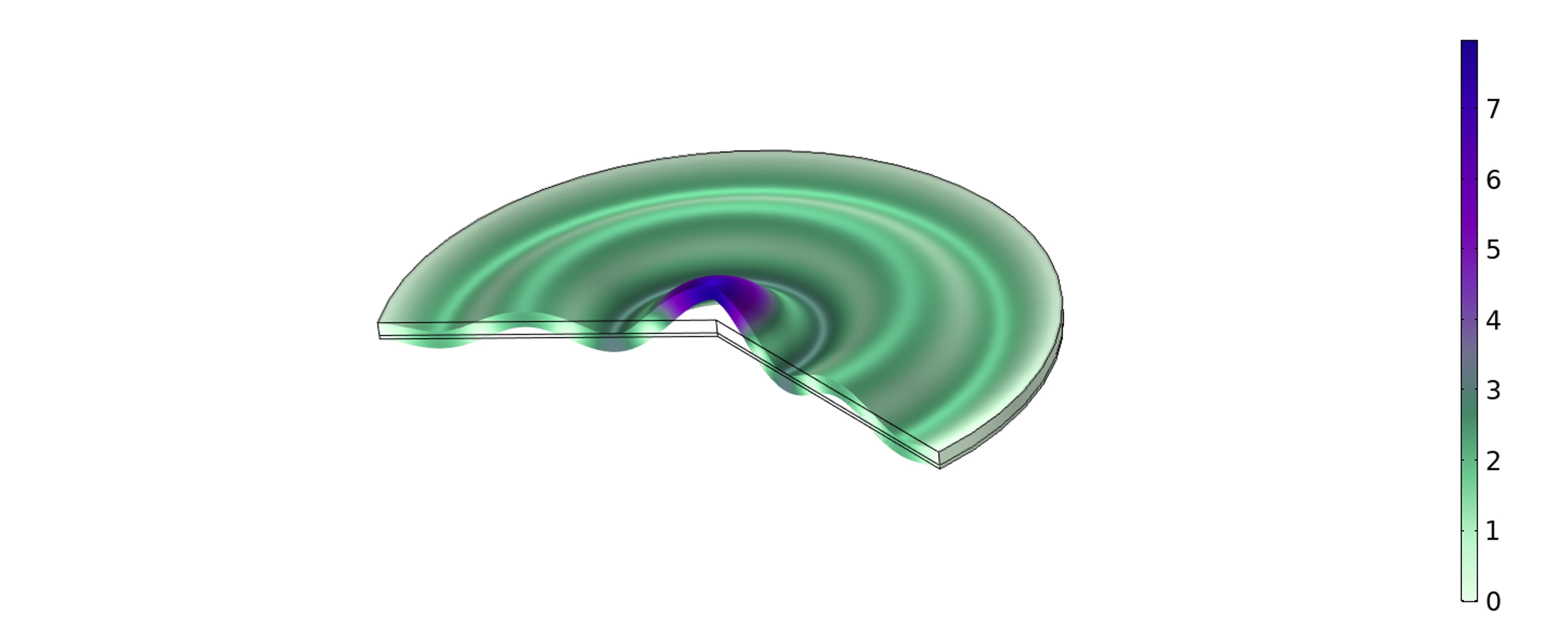} }}%
    \caption{Modal analysis of the circular plate for 4 respective resonance frequencies:
a) First resonance frequency: 9.74 kHz b) Second resonance frequency: 36.44
kHz c) Third resonance frequency: 77.30 kHz d) Fourth resonance frequency: 128.91
kHz.}\label{fig:fig5}
\end{center}\vs{-4mm}
\end{figure}

\begin{figure}[H]
\begin{center}
    \centering
    \subfloat[\centering]{{\includegraphics[width=7cm]{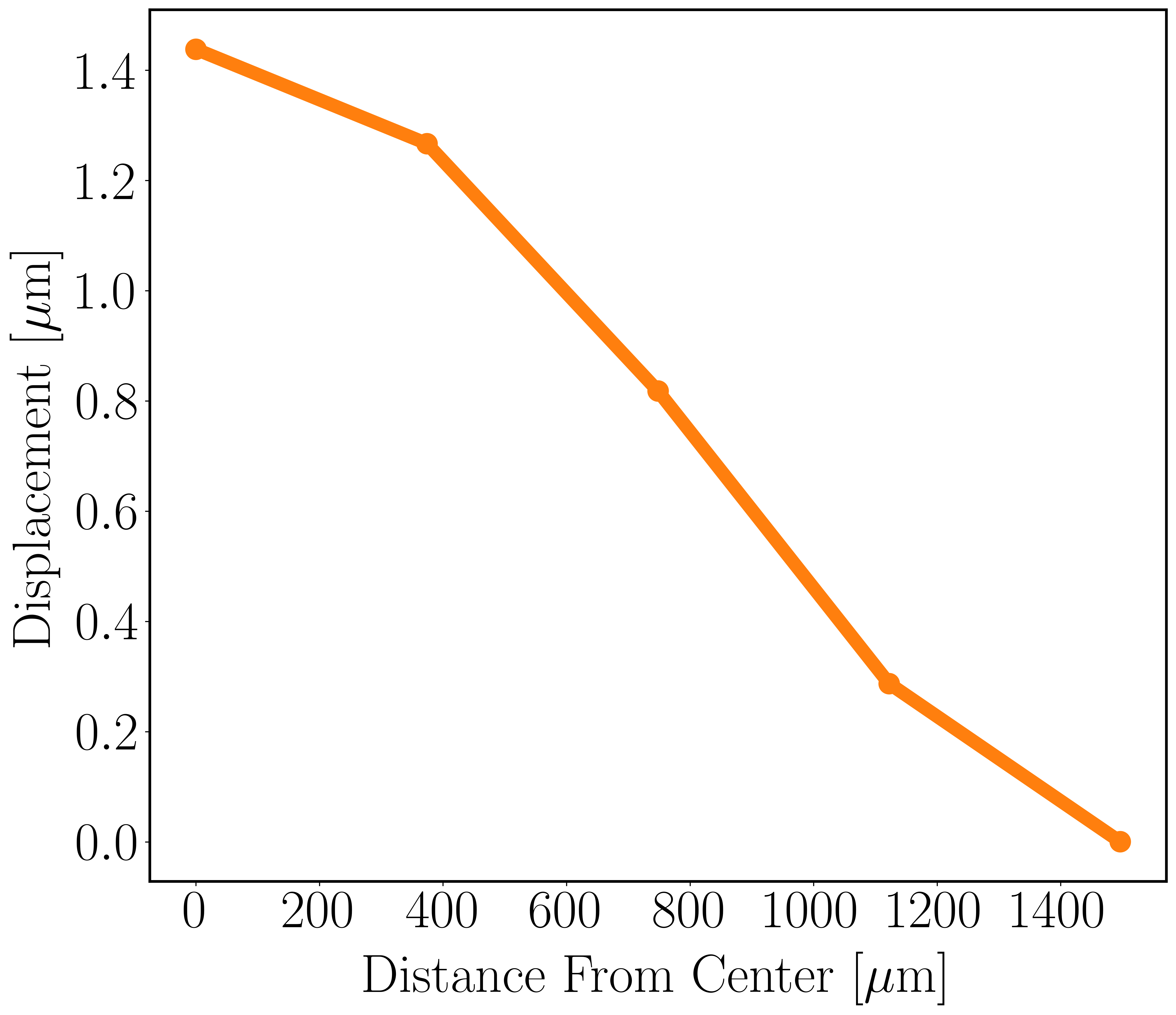} }}%
    \qquad
    \subfloat[\centering]{{\includegraphics[width=7cm]{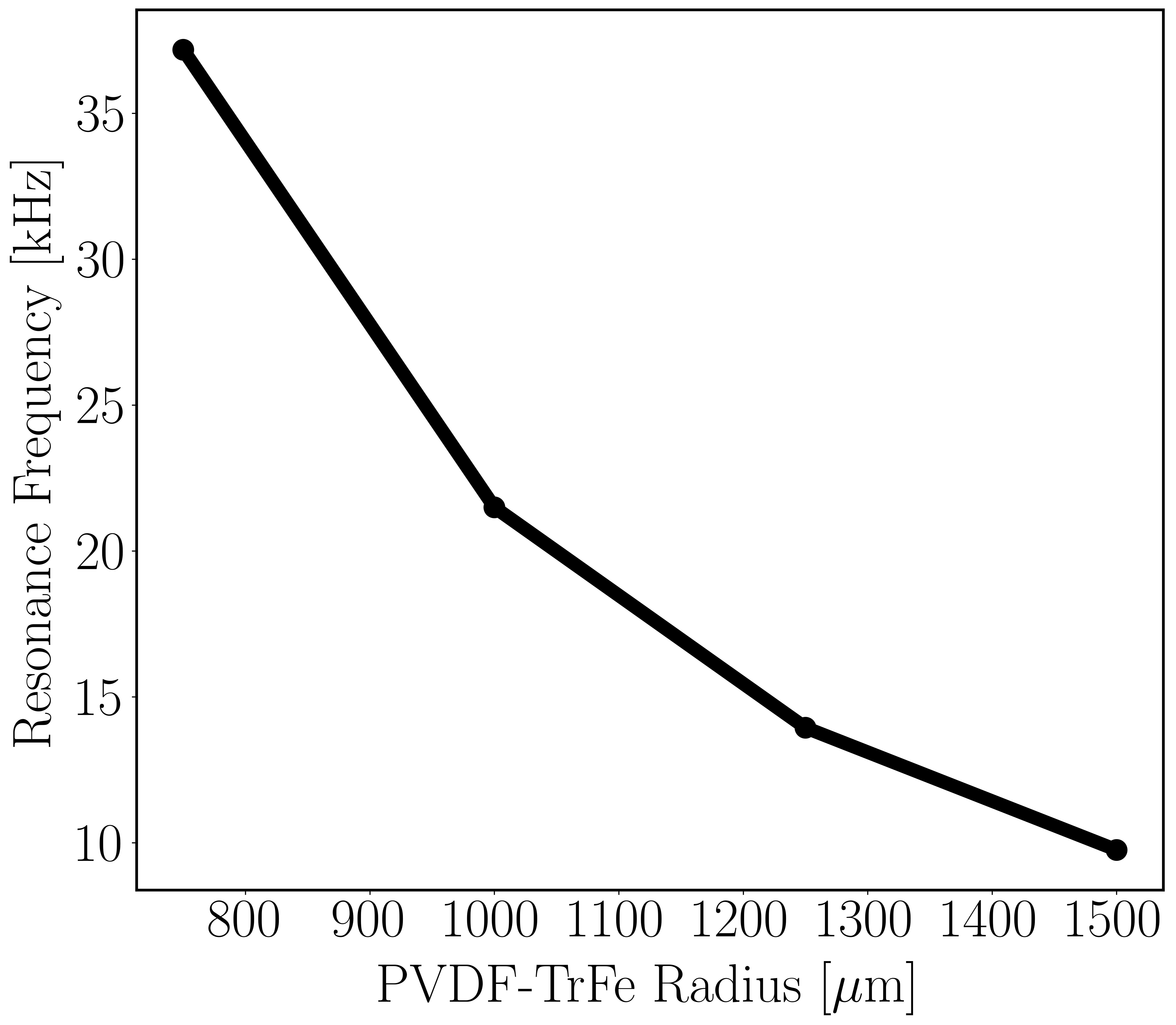} }}%
    \vskip\baselineskip
    \centering
    \subfloat[\centering]{{\includegraphics[width=9cm]{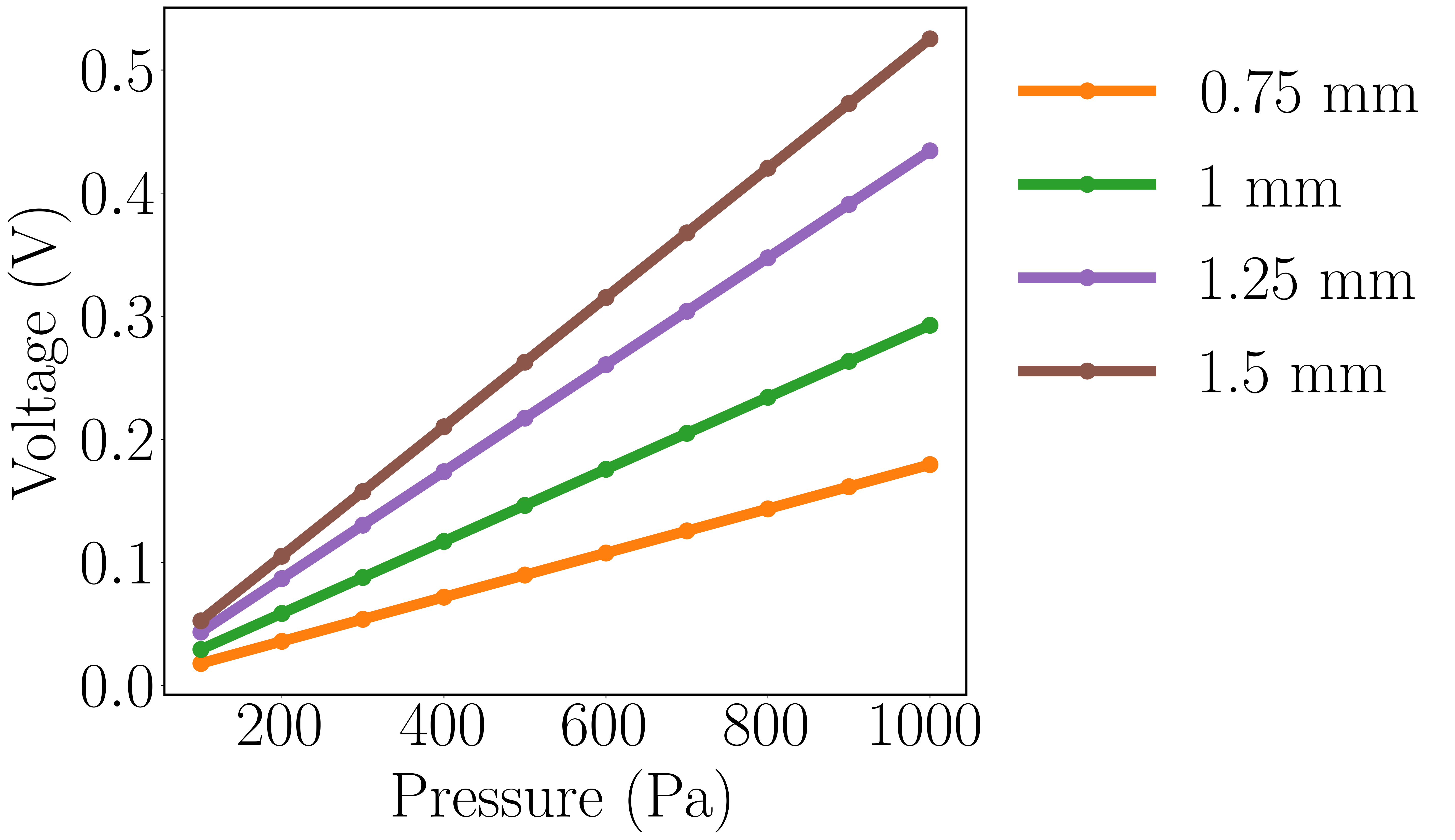} }}%
    \qquad
    \subfloat[\centering]{{\includegraphics[width=7cm]{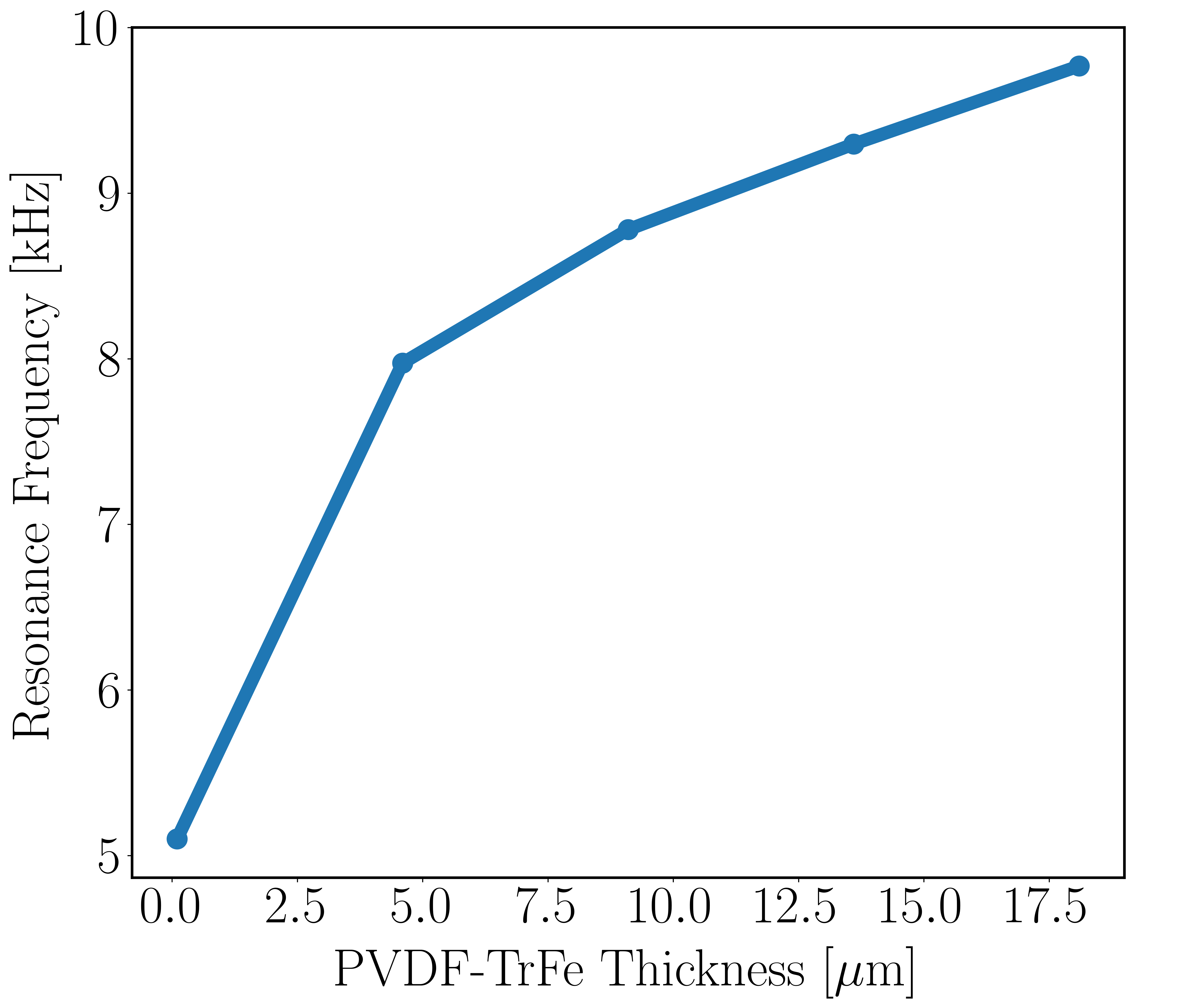} }}%
    \caption{Mechanical properties of the device for optimization: a) Displacement
of a selected point on the plate surface for increasing distance from the center b)
Resonance frequencies for increasing radius c) Voltage outputs of the circular plates for increasing pressure values d) Resonance frequencies for increasing thickness.}\label{fig:fig7}
\end{center}\vs{-4mm}
\end{figure}

\begin{figure}[H]
\begin{center}
    \centering
    \subfloat[\centering]{{\includegraphics[width=7cm]{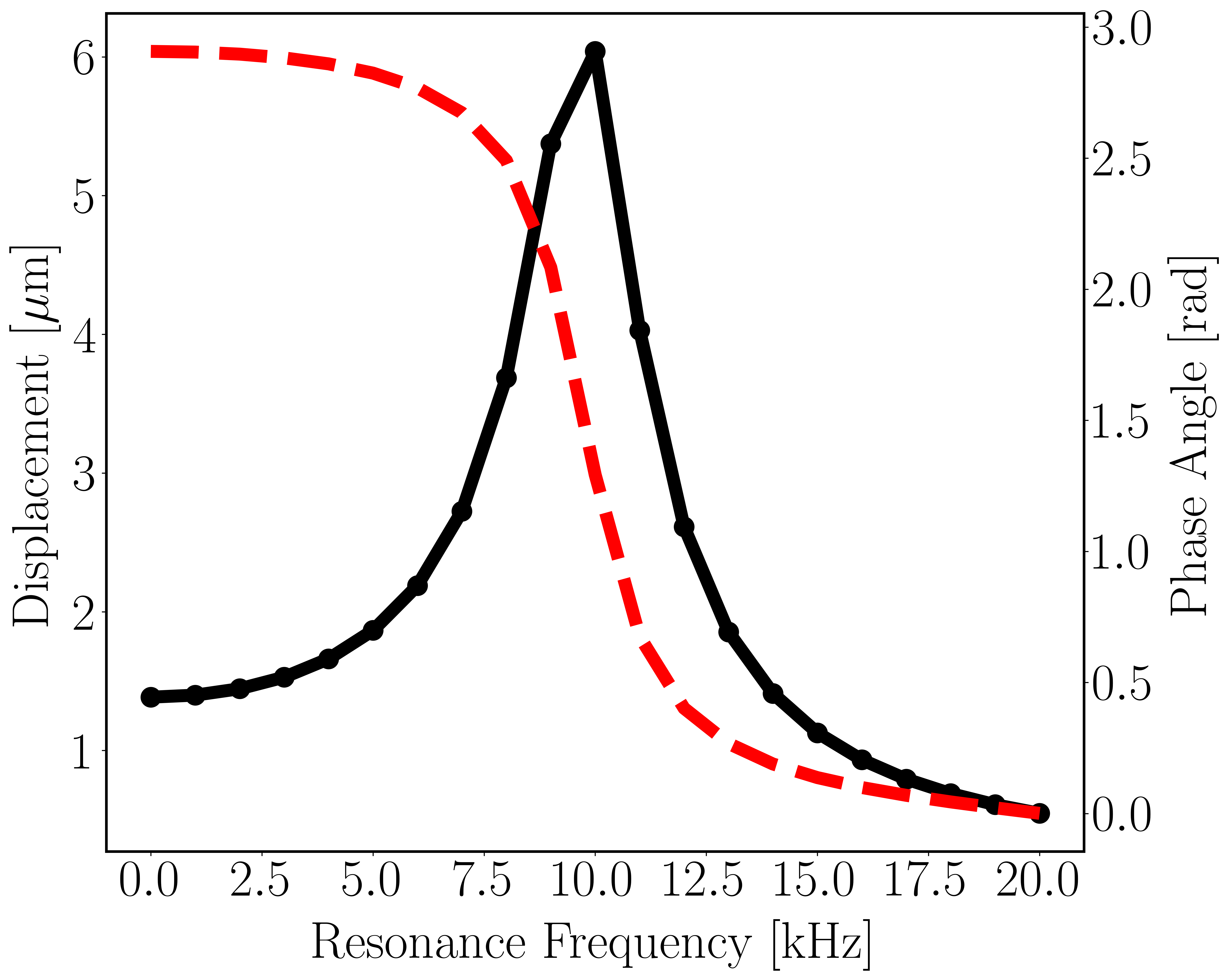} }}%
    \qquad
    \subfloat[\centering]{{\includegraphics[width=7cm]{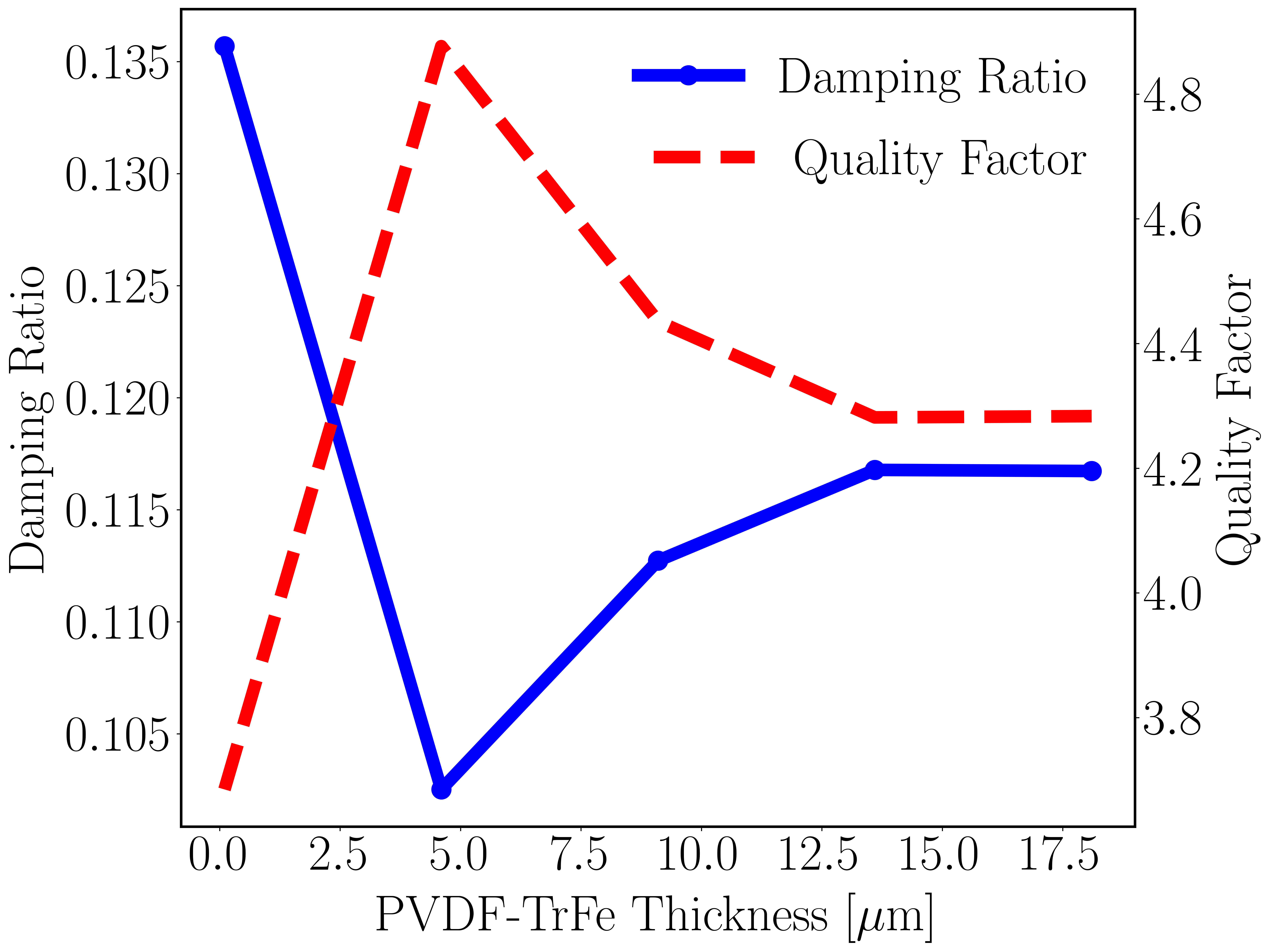} }}%
    \vskip\baselineskip
    \centering
    \subfloat[\centering]{{\includegraphics[width=7cm]{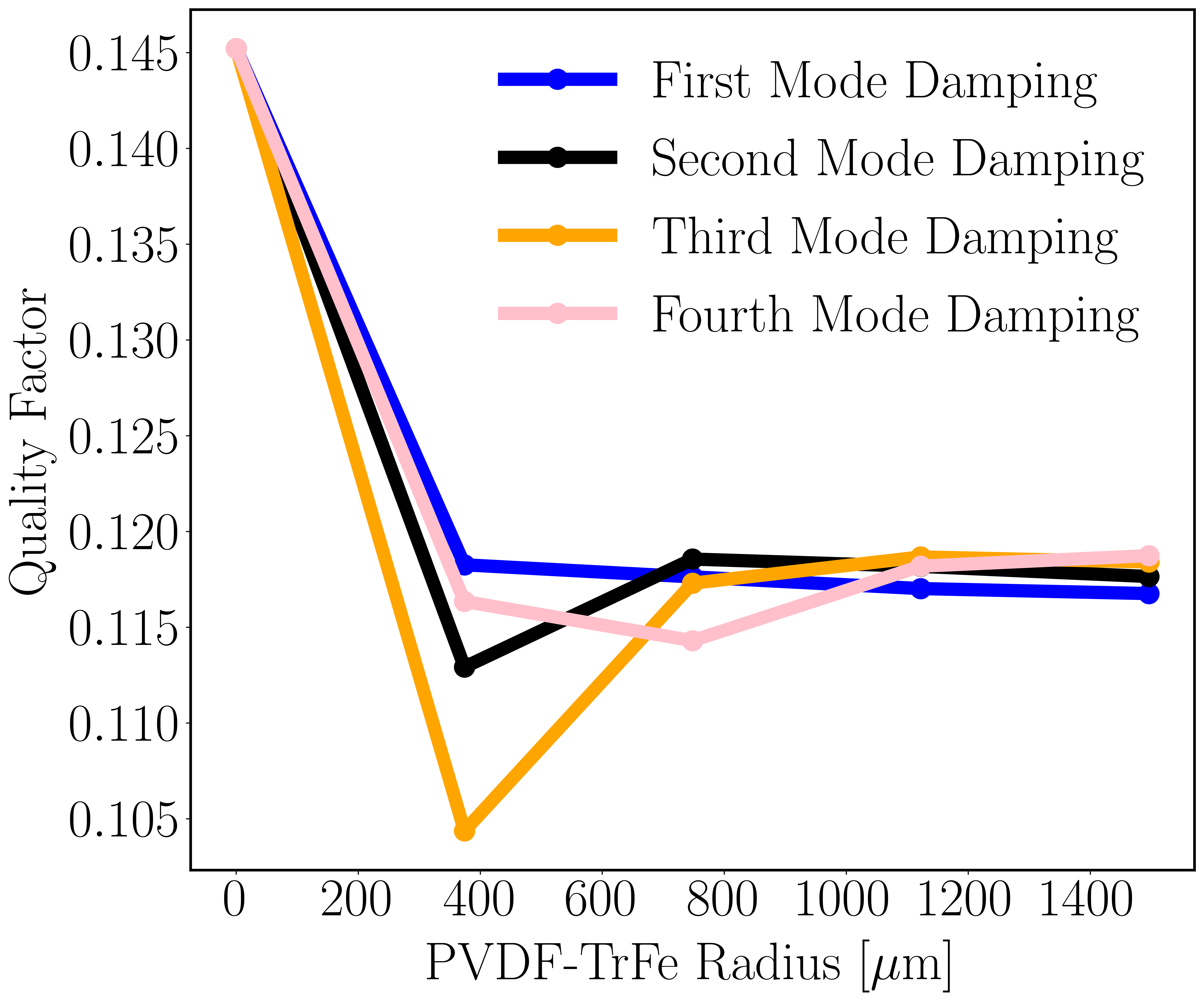} }}%
    \qquad
    \subfloat[\centering]{{\includegraphics[width=9cm]{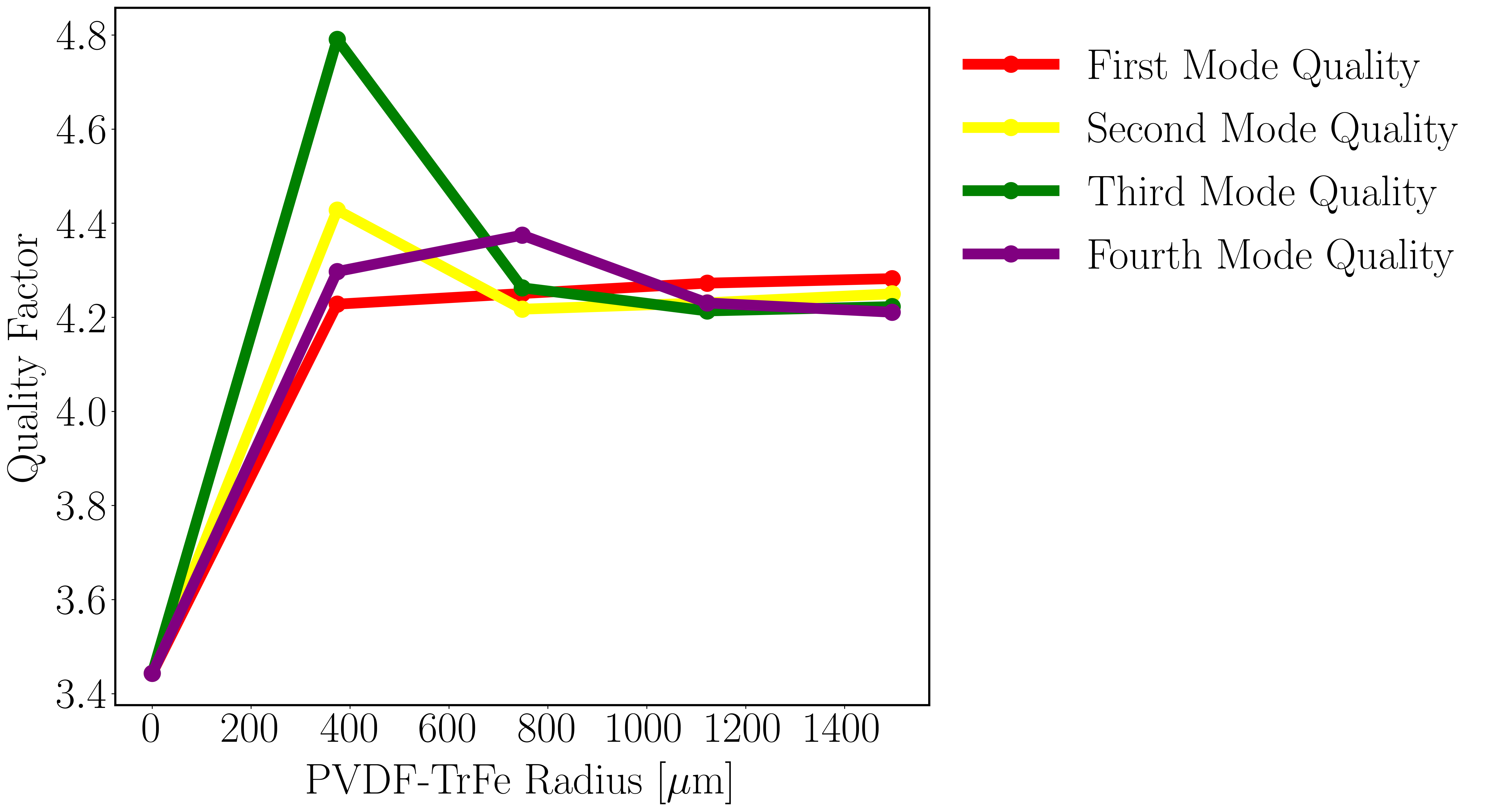} }}%
    \caption{Frequency response plots of the circular plate: a) Frequency response
of the plate for phase change and amplitude b) Damping ratio and quality factor of
the plate change for increasing piezoelectric layer thickness c) Damping factor changes for increasing plate radius d) Quality factor changes for increasing plate radius. }\label{fig:fig8}
\end{center}\vs{-4mm}
\end{figure}

To model the sensor, an electromechanical equivalent circuit was used. In the circuit, there is a capacitor in the electrical domain and spring and mass elements in the mechanical domain {[}33{]}.
\begin{equation}
\label{eq3}
C = {\varepsilon\pi{r_{pm}}^2}/{t_{pvdf}}.
\end{equation}
where $\varepsilon$ is the ratio of PVDF's relative permittivity to the permittivity of a vacuum, $t_{pvdf}$ is the thickness of PVDF, and $r_{pm}$ is the radius of the top electrode.With a normal plate deflection shape function of $\varphi$(x) {[}34{]}:
\begin{equation}
\label{eq4}
\varphi(x) = {(1-{x_\theta}^2)}^2
\end{equation}
where $x_\theta$ is the normalized version of the radial coordinate. The axisymmetric plate deflection is given by w($x_\theta$)=$w_0\varphi$(x), where $w_0$ is the static plate deflection at the center of the clamped plate. $I_m$, the piezoelectric coupling integral, and $I_e$, the strain energy integral, can be calculated using the formulas {[}33{]}:
\begin{equation}
\label{eq5}
I_m = 1/{(1-v)} \int_{0}^{\gamma} {((x_{\theta}d^{2}\varphi(x_\theta)}/{d{x_\theta}^2)}+{(d\varphi(x_\theta)}/{dx_\theta))dx_\theta}
\end{equation}
\begin{equation}
\label{eq6}
I_e = \int_{0}^{1}(( {x_{\theta}{{d^{2}\varphi(x_\theta)}/{d{x_\theta}^2})^2}+2v{({d\varphi(x_\theta)}/{x_{\theta}dx_{\theta})}{(d^{2}\varphi(x_\theta)}/{d{x_\theta}^2})}+{(d\varphi(x_\theta)}/{dx_\theta))}d{x_{\theta}}}
\end{equation}
where v is the effective Poisson's ratio of the composite plate and M is the piezoelectric bending moment supplied by the piezoelectric material [35]:
\begin{equation}
\label{eq7}
M = -e_{31,f}V_{in}z
\end{equation}
$e_{31,f}$ is the transverse piezoelectric coefficient. $V_{in}$ is the applied voltage. In addition, z is the distance between the midplane of the active PVDF layer and the neutral plane. Notably, the piezoelectric coupling integral $I_m$ is only calculated across the electrode region (normalized radial coordinate 0 to), whereas the strain energy integral $I_e$ is calculated throughout the entire radius. The location of the composite plate structure's neutral plane z, is determined by the following:
\begin{equation}
\label{eq8}
z = {\sum_{k=1}^{3} {(t_{k}z_{k}E_{k}}/{1-v_{k}^2)}}/{\sum_{k=1}^{3}(t_{k}E_{k}}/{1-v_{k}^2)}
\end{equation}
The subscripts represent the thin plate layers from the harvester's base in ascending order. $E_{k}$ is Young's Modulus, Poisson's ratio equals $v_{k}$, and each layer's central axis is $z_{k}$. Similarly, the flexural rigidity, D [36], and effective mass per unit area, $\mu_{eff}$ of the composite plate are:
\begin{equation}
\label{eq9}
z = \sum_{k=1}^{3} 1/3( {{{(h_{k}-z)}^3}-{(h_{k-1}-z)}^3)}/{({1-{v_{k}}^2}/E_k)}
\end{equation}
\begin{equation}
\label{eq10}
\mu_{eff}=(1/D)\sum_{k=1}^{3} \rho_{k}t_k
\end{equation}
$\rho_{k}$ represents the density, while $h_k$ represents the height of each layer.
When solving (\ref{eq13}) for the $1/k_m$ mechanical compliance, one obtains:
\begin{equation}
\label{eq11}
1/k_m = {r^2}/{2{\pi}DI_e}
\end{equation}
Solving gives the following equation for the electromechanical coupling ratio:
\begin{equation}
\label{eq12}
\eta = 2{\pi}I_{m}e_{31,f}z
\end{equation}
The circular plate formula [34] is used to determine the natural frequency:
\begin{equation}
\label{eq13}
f_{n}=({{{\lambda_{01}}/r)}^2}\sqrt{D/{\mu_{eff}}}
\end{equation}
where $\lambda_{01}$ is the vibration mode (01) eigenvalue. The mass of the composite disk as a whole, $m_d$. In addition, the shape function is utilized to calculate the modal mass:
\begin{equation}
\label{eq14}
m_{d}= ({\rho_{\rho\rho}t_{\rho\rho}+\rho_{be}t_{be}+\rho_{a\rho}t_{a\rho})\pi\\r^2\\+{\rho_{te}t_{te}\pi\\{r_{te}}^2}}
\end{equation}
\begin{equation}
\label{eq15}
m_{m}= m_{d}*2\int_{0}^{1} {{\varphi(x_\theta)}^2}x_{\theta}dx_{\theta}
\end{equation}
where the densities of the various layers are $\rho_{\rho\rho}, \rho_{be},\rho_{a\rho}$ and $\rho_{te}$ respectively. The Finite element modeling of the device was made using \textsc{COMSOL}$^{\circledR}$. Von Mises stress analysis is given in figure (\ref{fig:fig4}).
To visualize the modal frequencies of the circular plate, a FEM model for modal analysis is given in figure (\ref{fig:fig5}).
The voltage output plot of the circular plate is given in figure (\ref{fig:fig6}). The simulation was run for increasing radius values for 400 Pa pressure input.
For different parametric measurements, mechanical properties of the device for optimization are given in figure (\ref{fig:fig7}).
The frequency response of the circular plate is given in figure (\ref{fig:fig8}). Phase change, amplitude, damping ratio, and quality factor values are given.

\section{Results and Discussion}
\begin{figure}[h!]
\begin{center}
    \centering
    \subfloat[\centering]{{\includegraphics[width=7.5cm]{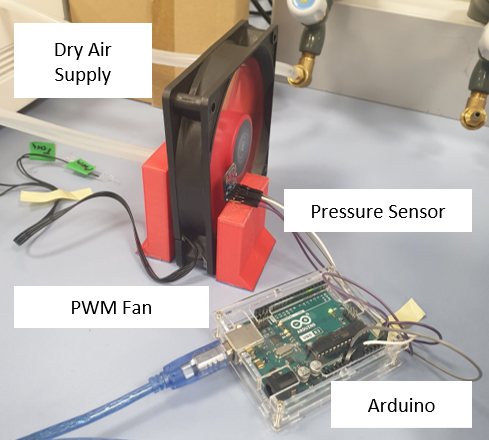} }}%
    \qquad
    \subfloat[\centering]{{\includegraphics[width=8cm]{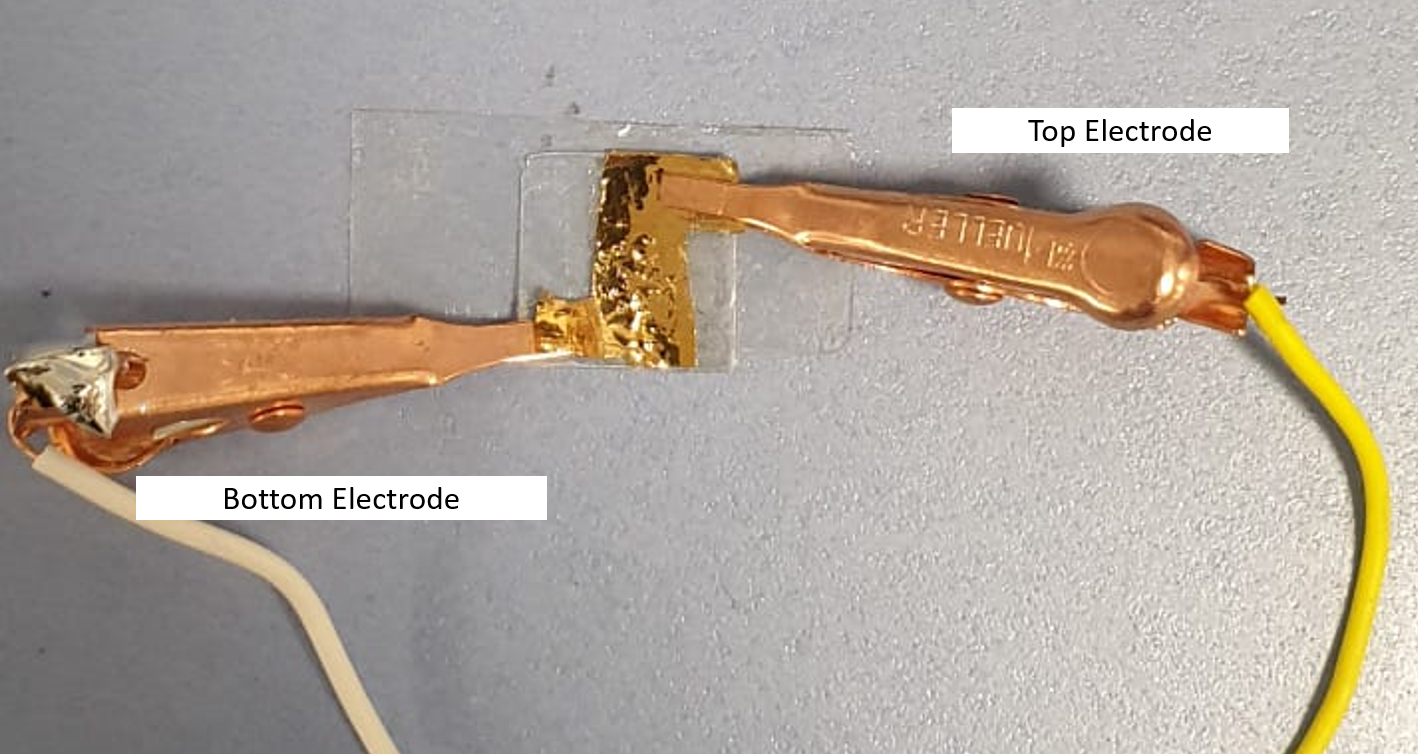} }}%
    \caption{The piezoelectric harvester is shown under test: a) 3D-printed parts with a cavity for dry air supply and a surface for mounting the pressure sensor were used to construct the test setup. SPI communication was utilized to obtain the results from the pressure sensor b) The piezoelectric element was run under test getting connections from both electrodes.}\label{fig:fig9}
    \end{center}\vs{-4mm}
\end{figure}

The experimental platform was constructed as depicted in figure (\ref{fig:fig9}) with the device under test . The configuration included a 4-wire Pulse Width Modulation (PWM) fan, an Arduino Uno, a dry air source, and an Adafruit BMP280 pressure sensor breakout board.

A code was built to sense pressure from serial pins dynamically. The Serial Peripheral Interface (SPI) protocol was utilized for communication. Using 3D-printed components, a novel test setup was built. Dry air was directed toward the fan's rotating propeller, which generates pressure fluctuations by generating intervals. The frequency of these intervals was determined by adjusting the fan's PWM frequency and operating voltage. PEH was attached close to the fan, so these variations could vibrate it to generate alternating voltage. The maximum wind speed recorded was 20 meters per second. The fan's rotational speed was confirmed using a tachometer.

\begin{figure}[H]
\begin{center}
    \centering
    \subfloat[\centering]{{\includegraphics[width=6cm]{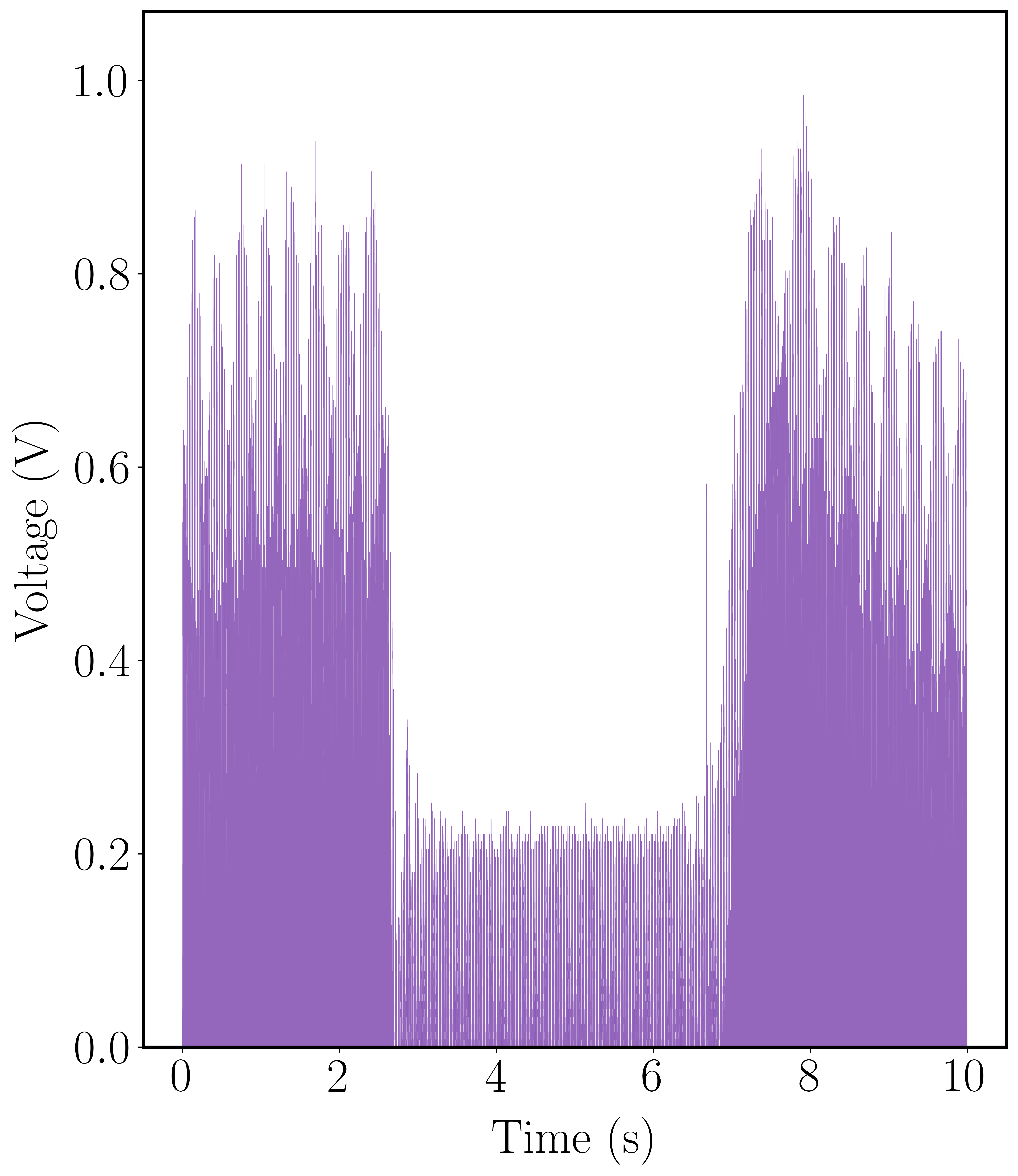} }}%
    \qquad
    \subfloat[\centering]{{\includegraphics[width=10cm]{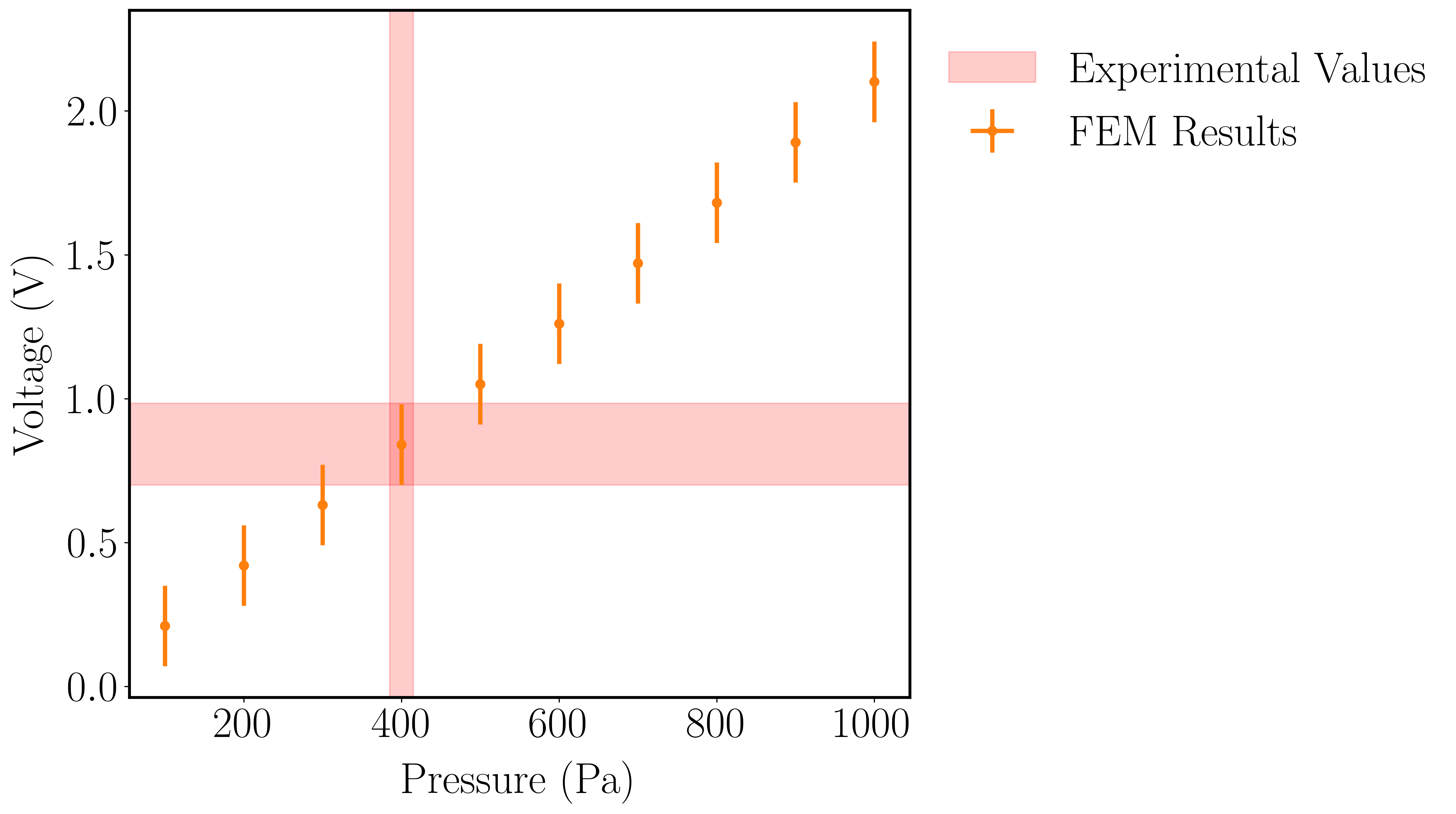} }}%
    \caption{Frequency response plots of the circular plate: a) $V_{max}$ was measured as 984 mV while $V_{pp}$ = 1.74 V was calculated from the wind tests b) Error bars represent a 10.1 $\%$ error between the experimental positive maximum voltage value of 756 mV and the FEM result for pressure application 400 ± 30 Pa}\label{fig10}
    \end{center}\vs{-4mm}
\end{figure}

Figure (\ref{fig10}) depicts the sensor's voltage output under wind load without any external circuit connection, and the comparison of FEM values against values taken from the wind setup for voltage.
\section{Conclusion}
In this study, mechanical and electrical characteristics of a piezoelectric wind energy harvester are given. The properties of the output voltage were examined using a wind test instrument. For von Mises stress and modal analysis, finite element modeling (FEM) was used. The circular plate's resonance frequency sweeps, quality factors, and damping ratios were given numerically. For a PVDF-TrFe piezoelectric layer with a thickness of 18 $mu$m and a radius of 1.5 mm, the damping ratio and quality factor were calculated to be 0.117 and 4.284, respectively. $V max$ was determined to be 984 mV based on the wind configuration and compared to the FEM outputs.
\section*{Acknowledgment}
B.K. did the experiments and wrote the paper. L.B. supervised the research and gave feedback for publication.

\end{document}